# Quantum-Chemistry based design of halobenzene derivatives with augmented affinities for the HIV-1 viral G$_4$/C$_{16}$ base-pair.

Perla El Darazi[1, 2], Léa El Khoury [1, 2, 3], Krystel El Hage[4], Richard G. Maroun[2], Zeina Hobaika[2],

Jean-Philip Piquemal*[1, 5,6], Nohad Gresh*[1]

1 Sorbonne Université, Laboratoire de Chimie Théorique, UMR7616 CNRS, Paris, France

2 UR EGP, Centre d'Analyses et de Recherche, Faculté des Sciences, Université Saint-Joseph de Beyrouth, Beirut, Lebanon

3 Present address: Department of Pharmaceutical Sciences, University of California, Irvine, Irvine, CA, USA

4 Université Paris-Saclay, Inserm, Univ Evry, Structure-Activité des Biomolécules Normales et Pathologiques, 91025, Evry, France

5 Institut Universitaire de France, Paris, France

6 Department of Biomedical Engineering, The University of Texas at Austin, Austin, TX, USA.

**Corresponding authors:**

<u>nohad.gresh@lct.jussieu.fr</u> **(NG),**

<u>jean-philip.piquemal@sorbonne-université.fr</u> **(JPP)**

**Abstract.**

The HIV-1 integrase (IN) is a major target for the design of novel anti-HIV inhibitors. Among these, three inhibitors which embody a halobenzene ring derivative (HR) in their structures are presently used in clinics. High-resolution X-ray crystallography of the complexes of the IN-viral DNA transient complex bound to each of the three inhibitors showed in all cases the HR ring to interact within a confined zone of the viral DNA, limited to the highly conserved 5'CpA 3'/5'TpG 3' step. The extension of its extracyclic CX bond is electron-depleted, owing to the existence of the 'sigma-hole'. It interacts favorably with the electron-rich rings of base $G_4$. We have sought to increase the affinity of HR derivatives for the $G_4/C_{16}$ base pair. We thus designed thirteen novel derivatives and computed their Quantum Chemistry (QC) intermolecular interaction energies (ΔE) with this base-pair. Most compounds had ΔE values significantly more favorable than those of the HR of the most potent halobenzene drug presently used in clinics, Dolutegravir. This should enable the improvement in a modular piece-wise fashion, the affinities of halogenated inhibitors for viral DNA (vDNA). In view of large scale polarizable molecular dynamics simulations on the entirety of the IN-vDNA-inhibitor complexes, validations of the SIBFA polarizable method are also reported, in which the evolution of each ΔE(SIBFA) contribution is compared to its QC counterpart along this series of derivatives.

**Introduction**

The HIV-1 integrase (IN) catalyzes the transfer of a viral DNA (vDNA) strand into the genome of the host cell (Lesbats et al., 2016). It is also involved in reverse transcription (Hironori et al, 2009), nuclear import (Mouscadet et al., 2007) and HIV-1 particle maturation (Kessl et al., 2016). It has no counterpart in human cells and thus constitutes an emerging target for the design of novel anti-retroviral inhibitors (Liao et al., 2010).

Three integrase inhibitors have been approved by the FDA in anti-HIV therapies, Raltegravir (RAL) (Summa et al., 2008), Elvitegravir (EVG) (Shimura et al., 2008) and Dolutegravir (DTG) (Underwood et al., 2012). All three act as integrase strand transfer inhibitors (INSTIs) (Ammar et al., 2016). They embody two distinct structural motives, namely a large diketo acid pharmacophore, and a halobenzene derivative. A major advance toward the design of novel derivatives is enabled by high-resolution X-ray crystal structures of the ternary complexes of IN, vDNA, and each of the three INSTI's (Hare et al., 2010; Hare et al., 2011). These show all three drug complexes to be two-pronged. The keto oxygen and a coplanar neighboring oxygen both coordinate two IN catalytic Mg (II) cations, structural water molecules, and, either directly or through water, IN residues. The halobenzyl moiety is confined in a narrow cleft, binding to the $G_4$ and $C_{16}$ bases of the highly conserved 5'CpA 3'/5'TpG 3' step on the viral DNA ends (Hare et al., 2010). The emergence of IN mutations weakening IN-drug interactions is a major limitation for INSTI-based therapies (Wainberg et al., 2011). Such mutations were reported to occur solely on IN and not on vDNA: this thus leaves open the possibility that additional enhancements of INSTI-vDNA binding should not be adversely impacted by IN mutations.

The present study was motivated by two findings. First, reports from one of our Laboratories showed the DTG > EVG > RAL ranking of affinities for the IN-vDNA complex (denoted as the intasome, INT) to be paralleled by their corresponding affinities for the sole vDNA (El Khoury et al., 2017). Could, thus, increases of the INSTI-INT binding affinities be attempted upon focusing on the sole 'ternary' complexes of $G_4$, $C_{16}$, and a halobenzene ring? Second, recent spectrometric and computational studies of the binding of INSTIs to viral DNA extremities showed the ranking of affinities to be governed by the enthalpy ($\Delta H$) component of the binding free energies ($\Delta G$), the entropy component ($T\Delta S$) being similar for all three complexes (El Khoury et al., 2019), a 'signature' for intercalative binding (Chairs et al., 2006). Furthermore, the ΔH ranking of the three inhibitor affinities was itself paralleled by the corresponding ranking of the ab initio quantum chemistry (QC) intermolecular interaction energies, ΔE(QC), of their halobenzene rings with the *sole* $G_4/C_{16}$ base pair. Upon focusing on the halobenzene ring of the best bound compound, namely DTG, could, then, ΔE (QC) Energy Decomposition Analysis (EDA) along with electronic structure considerations offer insight for affinity-enhancing chemical substitutions?

The full structures of RAL, EVG and DTG are presented in figure 1. There is a zone of electron depletion along the extension of the C-F bond of DTG, denoted as the 'sigma-hole' (Murray et al., 2008). This bond is *para* to the C-C bond connecting the ring to the central diketo acid group (Figure 1). In the crystal structure of the DTG-INT complex, it points towards the electron-rich bicyclic ring of $G_4$. This could constitute a key stabilizing feature of the DTG-$G_4$ complex.

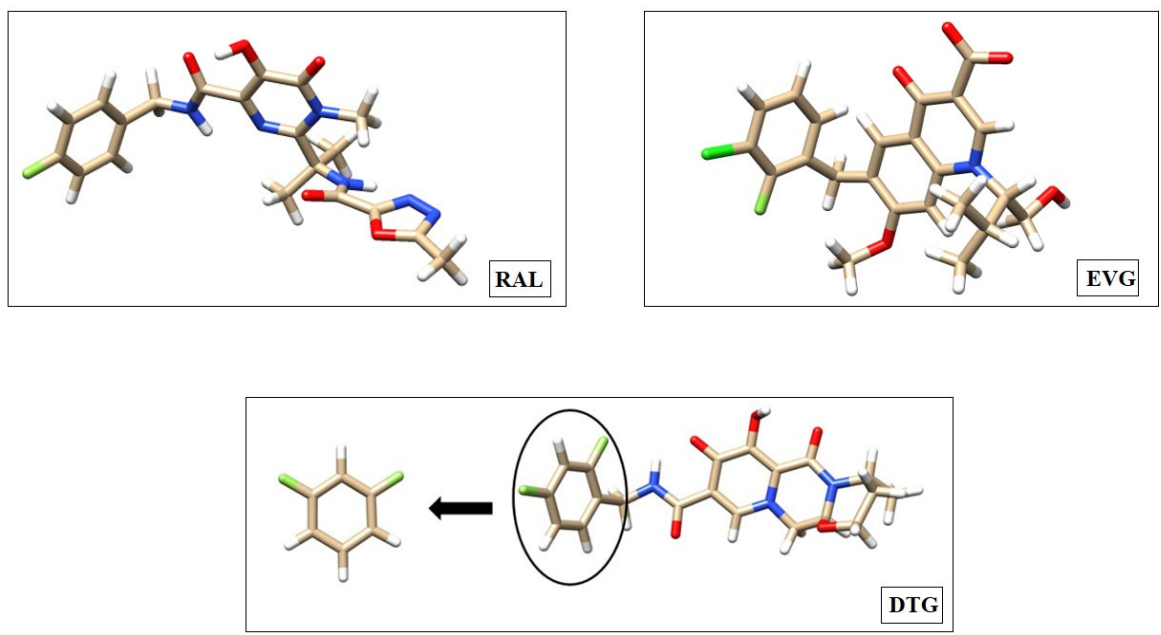

**Figure 1.** Molecular structure of Raltegravir (RAL), Elvitegravir (EVG) and Dolutegravir (DTG) represented with the Chimera software. The coordinates used for these representations are taken from the Protein Data Bank crystal structures of the Prototype Foamy Virus (PFV) intasome in complex with magnesium and Raltegravir (PDB code: 3OYA - Hare et al., 2010a), Elvitegravir (PDB code: 3L2U, Hare et al., 2010b) as well as with Dolutegravir (PBD code: 3S3M - Hare et al., 2011). We emphasized with a circle the difluoro halobenzyl ring of DTG, where the substitutions are going to be made to conceive new inhibitors. Carbon atoms are colored tan, hydrogen atoms are colored white, fluorine atoms are colored light green, chlorine atoms are colored dark green, nitrogen atoms are colored blue and oxygen atoms are colored red.

We have considered and analyzed several analogs of the DTG halobenzene ring (HR), thirteen of which will be reported in this paper. It is noted that with the exception of compound A1, all compounds have an extracyclic -NHCH$_3$ or -NH$_2$ proton donor replacing in *para* the extracyclic CF bond. All additional substitutions were done *ortho* or *para* to the C-C connector by electron-withdrawing groups. Each substituent can impact ΔE(QC) by a combination of different factors:

-on account of its electron-withdrawing character, a favorable increase of the electrostatic contribution, $E_C$, of ΔE (QC), of the *para* substituent with $G_4$;

-a favorable increase of the polarization contribution, $E_{pol}$, of ΔE (QC), due to the contribution of its polarizability to the polarization energy of the ligand;

-when in *meta*, it could contribute to additional electrostatic interactions with sites belonging to $G_4$ or $C_{16}$, as reflected by $E_C$ and $E_{pol}$. There is a cone of electron-rich density around the CX bonds of halogens. Halogen substituents could enhance $E_C$ if such a cone was in the vicinity of electron-deficient sites of $C_{16}$.

-these three factors could be counteracted in some cases by increases of the short-range repulsion, $E_X$, and, in the energy balances, by a larger solvation penalty than DTG, since most designed derivatives bear a more polar character than DTG.

EDA unravels the relative magnitudes of the individual $\Delta E$ (QC) contributions along the series investigated, and how each contribution can be impacted by each substituent. Optimizing the localized HR-$G_4$-$C_{16}$ interactions could contribute to a modular design of INSTIs as a preliminary to long-duration molecular dynamics (MD) of the entirety of the INSTI with the entirety of the INT. The binding site includes highly polar protein residues, two Mg (II) cations, and structural, highly polarizable water molecules. Polarizable, multipolar Molecular Mechanics/Dynamics approaches, such as SIBFA (Gresh et al., 2007) or AMOEBA (Ponder et al., 2010) should be adapted to these simulations. Such procedures were demonstrated to reliably account for the impact of the sigma hole and the dual character of the CX bond of HR on electrostatics (El Hage et al.; 2013). Very large macromolecular complexes, such as the INT-INSTI ones, are now amenable to long-duration MD upon resorting to the massively parallel code Tinker-HP (Lagardère et al., 2018). Prior to these, and in the context of the present study, it was thus essential to evaluate the accuracy of one of these procedures, SIBFA. Specifically, how well will the evolutions of $\Delta E$(QC) and of each of its contributions, be paralleled by the SIBFA contributions along the series of the thirteen HR ligands?

As a complement to EDAs, we will also report the contours of their electrostatic potential maps.

**Methods**

**1. Quantum chemistry calculations**

*a. Energy decomposition analysis.*

The decomposition of the ab initio SCF interaction energy is done using the Reduced Variational Space (RVS) analysis (Stevens et al., 1987), where the intermolecular interaction energy is separated into four contributions: Coulomb ($E_C$) and short-range exchange-repulsion ($E_X$) in first order ($E_1$) and polarization ($E_{pol}$) and charge-transfer ($E_{ct}$) in second order ($E_2$). Finally, the dispersion contribution is assessed as the difference between the BSSE-corrected B97-D3 intermolecular interaction energies and the Hartree-Fock (HF) ones. The basis set superposition error (BSSE) is taken into consideration in the final energy values. The GAMESS software with the cc-pVTZ (-f) basis set (Schmidt et al., 1993) were used in this analysis. We denote below as $E_{pol}$ (VR), a 'variational' value of $E_{pol}$, obtained as the difference between ΔE(RVS) and the sum of $E_1$ and $E_{ct}$. This is to be contrasted to $E_{pol}$(RVS) at the outcome of the RVS procedure, the sum of all individual ligand polarization energies computed in a process when the occupied molecular orbitals (MO) of this ligand is relaxed towards its own virtual MO's, the other ligands being frozen.

Energy decomposition analyses were also performed at the correlated levels and the ω-B97-D3 functional (Grimme., 2006) resorting to the Absolutely Localized Molecular Orbitals method (ALMOEDA) (Azar et al., 2011) using the Q-Chem software (Shao et al., 2015). ΔE is decomposed into a 'frozen density' component (FRZ), namely the sum of the Coulomb and short-range contribution, and a polarization ($E_{pol}$) and a charge transfer contribution ($E_{ct}$) in second order.

*b. Correlated calculations.*

The intermolecular interaction energies (ΔE) of the complexes formed by the substituted rings with $G_4$ and $C_{16}$ were computed at the correlated level using the dispersion-corrected functionals B97-D3 and B3LYP-D3 (Goerigk et al., 2011), with the cc-pVTZ basis set (-f) (Feller., 1996) and the Gaussian software G09 (Frisch et al., 2009) software. The obtained values were corrected for BSSE (Simon et al., 1996). The values of the corresponding correlation and dispersion contributions were computed as the differences between the dispersion-corrected correlated ΔE and uncorrelated Hartree-Fock (HF) ΔE values.

G09 was used as well for energy-minimization of the complexes, using a starting structure determined by X-ray crystallography and taken from Protein Data Base site (PDB code: 3S3M, Hare et al., 2011). The C-C bond connecting the HR to the diketoacid (DKA) ring was replaced by a CH bond, since the DKA ring was removed.

The halobenzene rings were relaxed except for the hydrogen atom of this CH bond. This choice was made to avoid wanderings over the G4/C16 bases prevented in the complete complexes by HR anchoring to the rest of the drug. We also chose to hold both guanine and cytosine frozen in their X-ray geometry to account for their anchoring in the DNA backbone. We did not allow for conformational relaxation around the glycosidic bonds of G4 and C16 since we considered that this viral DNA base pair is held in an experimental position "tailored" for DTG: any additional positional relaxation in complexes with derivatives with bulkier ligands could *a priori* be expected to further optimize, rather than penalize, the binding of such derivatives, with possibly an even more favorable outcome than the one from the present study. This could clearly, only be clarified at the outcome of long-duration MD simulations on the ternary DNA-IN-ligand complexes, enabling to explore the widths and flexibility of the energy basins in the $G_4/C_{16}$/HR zone. Such studies will be reported in due course.

A Continuum desolvation energy $\Delta G_{solv}$, computed following the Polarizable Continuum Model (PCM) procedure (Mennucci et al., 2002) was also calculated for each HR ring. It was considered as an upper bound to its actual PCM desolvation energy prior to its complexation.

## 2. SIBFA computations

In the context of the SIBFA procedure, the intermolecular interaction energy ($\Delta E_{tot}$) is computed as the sum of five contributions: electrostatic multipolar ($E_{MTP}$), short-range repulsion ($E_{rep}$), polarization ($E_{pol}$), charge transfer ($E_{CT}$), and dispersion ($E_{disp}$). $E_{MTP}$ is computed with distributed multipoles (up to quadrupoles) derived from the QC molecular orbitals precomputed for each individual molecule, derived from the Stone analysis (Stone et al, 1985) and distributed on the atoms using a procedure developed by Vigné-Maeder and Claverie (Vigné-Maeder et al., 1988). $E_{MTP}$ is augmented with a penetration term (Piquemal et al., 2007). The anisotropic polarizabilities are distributed on the centroids of the localized orbitals (heteroatom lone pairs and bond barycenters) using a procedure due to Garmer and Stevens (Garmer et al., 1989). $E_{rep}$ and $E_{CT}$, the two short-range contributions, are computed using representations of the molecular orbitals localized on the chemical bonds and on localized lone-pairs. $E_{disp}$ is computed as an expansion into $1/R^6$, $1/R^8$, and $1/R^{10}$ and embodies an explicit exchange– dispersion term (Creuzet et al., 1991).

### 3. Contours of electrostatic potentials

The contours of molecular electrostatic potentials (MEPs) of the HRs derived from their electronic densities were displayed by the Gaussview software, implemented in the Gaussian software (Frisch et al., 2009). The colored zones are based on their electronic densities.

**Results and Discussion**

All DTG derivatives have at least one extracyclic halogen atom; F, Cl, or Br. The DTG coordinates are derived from the crystal structure of the viral intasome with Dolutegravir and two magnesium ions (PDB code: 3S3M, Hare et al., 2011). The first derivative considered, A1, has a chlorine substituent replacing the second DTG fluorine, which is *para* to the CH bond connecting it to the diketo moiety. This did not bring a significant $\Delta E_{tot}$ increase with respect to DTG (see Table 1 below). Several alternative substituents could be considered. However for the present study we deemed it more instructive to replace upfront the *para* -F group, which binds $G_4$ by its sigma-hole prolonging the CF bond, by another electron-deficient group, namely a proton donor. We could thus leverage these C-X bonds (X halogen or electron-deficient group) interactions and increase their magnitudes by depleting this donor electron-density with selected electron-attracting substituents. A natural choice bore on substituents such as -NHCH$_3$ or –NH$_2$. Both groups have the additional advantage of acting as electron-donating substituents: this is in contrast to fluorine, which is electron-withdrawing. This could favor the interaction of the electron-deficient sites of $C_{16}$ with the halobenzene ring and/or with the electron-rich cone around the halogen substituent. It is another means of leveraging the 'Janus-like' character of the halobenzene derivatives (El Hage et al., 2014, 2015). The –NHCH$_3$ substituted derivatives were split into two groups, having two or one electron-withdrawing substituents, respectively. The search for –NH$_2$ substituted derivatives was limited to three derivatives, with one or two fluorine substituents. The four groups are indexed as Group 1, 2, 3, and 4, and the derivatives in each group are indexed with capital letters from A to E. All derivatives are represented in Figure 2.

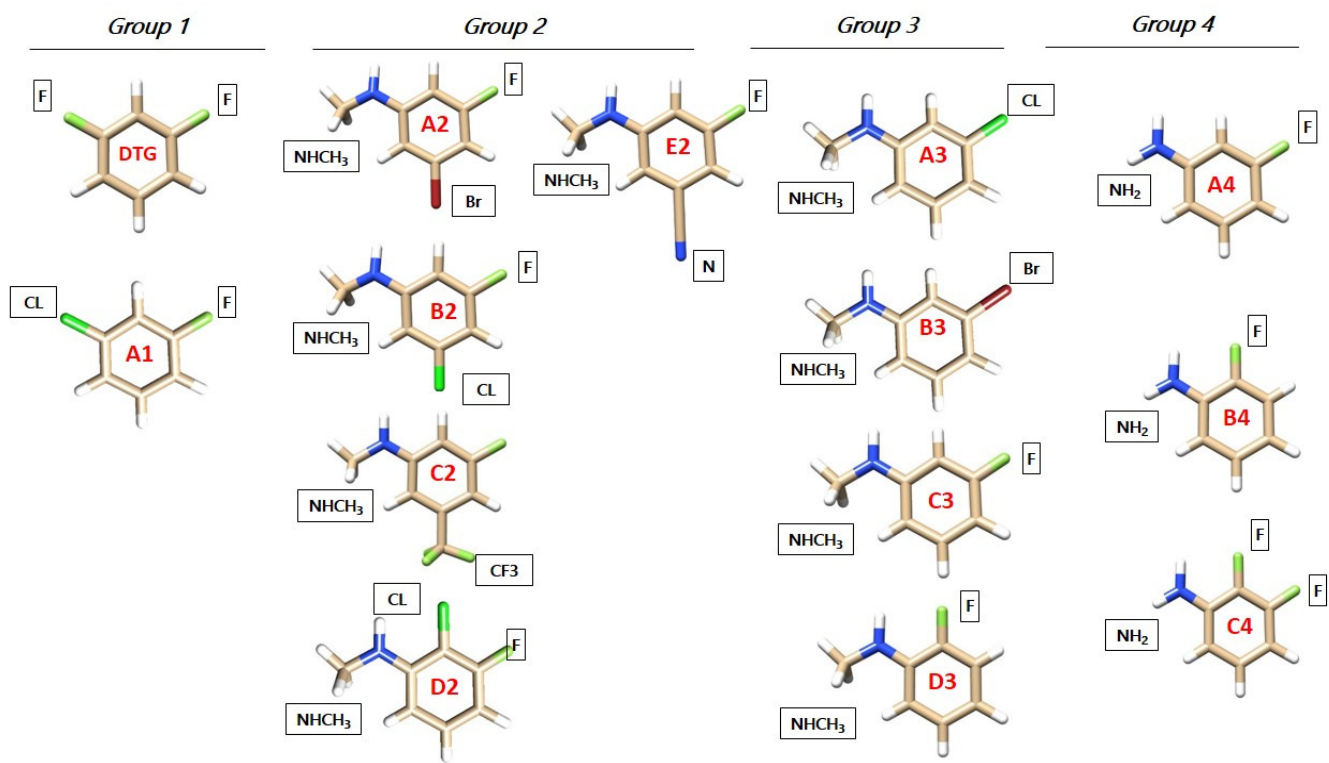

**Figure 2**. Molecular structures of the halobenzene ring of DTG (Group 1) and the newly designed ones (Groups 1, 2, 3 and 4) represented with the Chimera software. Carbon atoms are colored tan, hydrogen atoms are colored white, fluorine atoms are colored light green, chlorine atoms are colored dark green, bromine atoms are colored dark red and nitrogen atoms are colored blue. The squares indicate the new atoms or moieties chosen after substitution

*First group. Para-Electron Attracting substituted derivatives.* This group contains the compound A1 which has a similar structure to DTG, but with a chlorine atom in *para* instead of a fluorine one.

*Second group. Para–NHCH$_3$ substituted derivatives*. Five derivatives are considered. They all have, similar to DTG, fluorine in *ortho*, and a third, electron-withdrawing substituent:

A2, B2, and C2 have in the second *ortho* position: Br, Cl, and –CF$_3$, respectively. D2 has chlorine in *meta*, and E2 has a cyano substituent in the second *ortho* position.

*Third group. Para–NHCH$_3$ substituted derivatives.* A3, B3, and C3 have as *ortho* substituent chlorine, bromine and fluorine respectively, while D3 has fluorine in *meta* instead of *ortho*.

*Fourth group. Para–NH$_2$ substituted derivatives.* A4 and B4 have a fluorine in *ortho* and in *meta* respectively, while C4 has the two fluorines in *ortho* and *meta*.

Figure 3 gives a representation of the complexes of the G$_4$/C$_{16}$ vDNA base pair with representative compounds of series 1 to 4. We have displayed the parent compound, DTG from the first group, C2 and E2 from the second, B3 from the third and C4 from the fourth group. As seen below, the inhibitory rings face the ring of C$_{16}$, interacting with the latter via π-π stacking, while the moiety in *para* position face the electron-rich rings of G$_4$. Therefore, the electron donor groups in C2, E2, B3 and C4 as well as the electron-depleted region in the prolongation of the sigma-hole in DTG interacts with G$_4$ via electrostatic interactions.

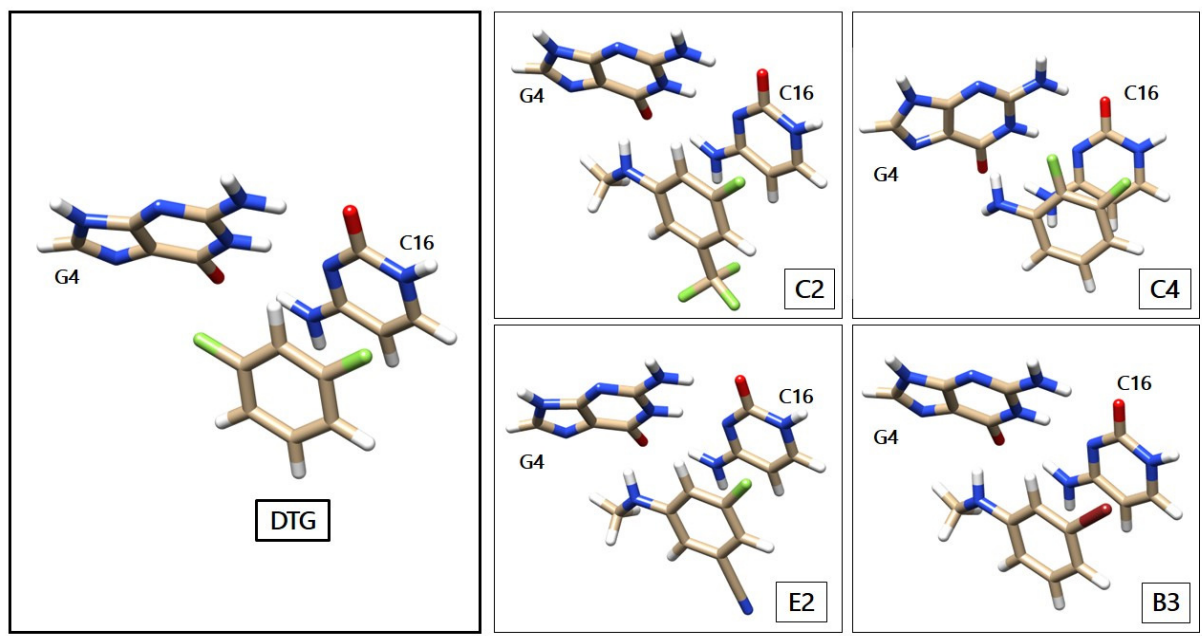

**Figure 3.** Representation of the $G_4/C_{16}$ complexes with representative HRs. The structures of $G_4$, $C_{16}$ and DTG are taken from the Protein Data Bank crystal structure of the Prototype Foamy Virus (PFV) intasome in complex with magnesium and Dolutegravir (PDB code: 3S3M Hare et al., 2011). The newly designed compounds are represented with the Chimera software. The color code for the atoms remains the same as in Figure 1, with the oxygen atoms colored in light red.

Table 1 lists the ΔE (QC) values at the B97-D3, B3LYP-D3 and ω-B97D levels together with the ΔE (SIBFA) values. It also lists the continuum solvation energies, $\Delta G_{solv}$, of each isolated derivative prior to complexation, and the resulting values of partial B97-D3 energy balances after $\Delta G_{solv}$ subtraction from ΔE (B97-D3). These $\Delta G_{solv}$ values should be considered as upper bounds to the actual PCM solvation energies of the halobenzene ring of DTG and its derivatives, because in the 'complete' inhibitor, the exposure of the polar group is less on account of a lesser accessibility and possible intramolecular interactions. The MD calculation of the actual solvation free energies of DTG and its derivatives in a box of water molecules will be considered in the next step of our studies in parallel with the MD simulations of their complexes with the intasome (El Darazi et al., work in progress).

The listed QC ΔE values take into account the BSSE correction as the HF and at the correlated levels. It has values in the 1.5-2 kcal/mol range at the HF level, and 3.5-4.6 kcal/mol range at the correlated levels, depending upon the functional, but changing little for a given functional with the ligand.

**Table 1**. Intermolecular interaction energies (kcal/mol) of halogenated derivatives with the $G_4/C_{16}$ vDNA base pair. DTG as well as the compounds displaying the best energetic values are highlighted in red.

| Compounds | ΔE B97-D3 (raw value/BSSE corrected) | ΔE B3LYP-D3 (raw value/BSSE corrected) | ΔE ALMOEDA (B97D3) | ΔE ALMOEDA (ω–B97D) | SIBFA | ΔG solv | Total interaction energy (ΔE B97-D3 -ΔG solv) (raw value/BSSE corrected) |
|---|---|---|---|---|---|---|---|
| **DTG** | **-37.4 / -33.2** | **-40.6 / -36.5** | **-33.2** | **-35.9** | **-39.6** | **-6.1** | **-31.3 / -27.1** |
| A1 | -37.5 / -33.8 | -40.6 / -37.0 | -33.8 | -35.8 | -38.4 | -5.9 | -31.6 / -27.9 |
| **A2** | **-45.1 / -40.7** | **-48.1 / -43.7** | **-40.6** | **-43.3** | **-45.9** | **-9.7** | **-35.4 / -31.0** |
| **B2** | **-44.9 / -40.5** | **-47.9 / -43.5** | **-40.4** | **-43.1** | **-46.6** | **-9.5** | **-35.4 / -31.0** |
| **C2** | **-46.0 / -41.3** | **-49.1 / -44.5** | **-41.2** | **-44.1** | **-47.3** | **-10.1** | **-35.9 / -31.2** |
| D2 | -43.0 / -38.6 | -46.3 / -41.9 | -38.5 | -41.0 | -45.2 | -7.8 | -35.2 / -30.8 |
| **E2** | **-45.3 / -40.9** | **-48.2 / -43.9** | **-40.6** | **-43.4** | **-47.7** | **-7.8** | **-37.5 / -33.1** |
| A3 | -44.7 / -40.4 | -47.6 / -43.4 | -40.3 | -42.7 | -44.6 | -9.4 | -35.3 / -31.0 |
| **B3** | **-44.9 / -40.7** | **-47.9 / -43.7** | **-40.6** | **-43.1** | **-46.9** | **-9.5** | **-35.4 / -31.2** |
| C3 | -44.1 / -39.6 | -47.0 / -42.6 | -38.8 | -41.5 | -45.2 | -9.7 | -34.3 / -29.9 |
| D3 | -42.6 / -37.8 | -45.7 / -41.1 | -38.4 | -41.2 | -44.0 | -7.8 | -34.8 / -30.0 |
| A4 | -42.5 / -38.2 | -45.2 / -40.9 | -38.0 | -40.6 | -43.7 | -11.2 | -31.3 / -27.0 |
| B4 | -41.5 / -37.0 | -44.2 / -39.7 | -37.4 | -39.7 | -41.1 | -9.7 | -31.8 / -27.3 |
| C4 | -42.1 / -37.5 | -45.0 / -40.4 | -41.2/ | -41.0 | -42.6 | -10.3 | -31.9 / -27.2 |

The evolution of ΔE as a function of the compound number is plotted in Figure 3 for each functional.

All three ΔE (QC) curves run parallel. Table 1 and Figure 4 show that the ΔE (QC) values of all compounds in series 2-4 have significantly larger magnitudes than those of either DTG or its chlorinated derivative A1. Taking the B97-D3 results as an example, the differences are in the 4-8.3 kcal/mol range. All three QC procedures concur into having five derivatives which stand out from the rest. Four belong to series 2, namely A2, B2, C2 and E2, and one belongs to series 3, namely B3. In series 2, all four derivatives have their two substituents on both *ortho* sides of the connecting -CH bond, or, equivalently, on both sides *meta* to the –NHCH$_3$ para substituent. The two best-bound compounds are C2 and E2, with either a –CF$_3$ or a cyano substituent –CN in *meta* to the –NHCH$_3$ group, respectively. C2 has a larger $\Delta G_{solv}$ value than E2, namely -10.1 compared to -7.8 kcal/mol. As mentioned above, this 2.3 kcal/mol difference should represent an upper bound to the actual desolvation energy difference between the two derivatives. It could reflect a greater 'hydrophilicity' of the C2 than the E2 ring. Including such a difference in partial relative energy balances would give rise to a 2 kcal/mol preference for the cyano derivative compared to the trifluoromethyl one.

In series 3, the best-bound derivative is B3, with a bromine substituent. Owing to the large hydrophilic character of this atom compared to fluorine in DTG, the values of $\Delta G_{solv}$ are correspondingly larger in series 2 and 3 than that of DTG, in the range 2.8-4.0 kcal/mol. This leaves out relative energy preferences for A2-C2, E2, and B3 over DTG in the range 4-6.7 kcal/mol.

The $\Delta E$ values in series 4 with an –NH$_2$ para substituent have smaller magnitudes than in series 2 and 3. These are 4-5 kcal/mol more favorable than DTG, but these are virtually fully compensated for by correspondingly larger $\Delta G_{solv}$ values, leaving partial energy balances less than 0.5 kcal/mol more favorable than DTG, which is inconclusive in terms of augmented affinities.

As for the BSSE values, we wish to note that the average values of the BSSE amounts to 4.14 kcal/mol for the B97D-3 functional and to 3.41kcal/mol for the W97D functional, both used for the ALMOEDA energetic calculations. With each, the relative variations among the compounds are around 0.5 kcal/mol. Whereas, at the HF level, the average value of BSSE is in the range of 1.5-2 kcal/mol.

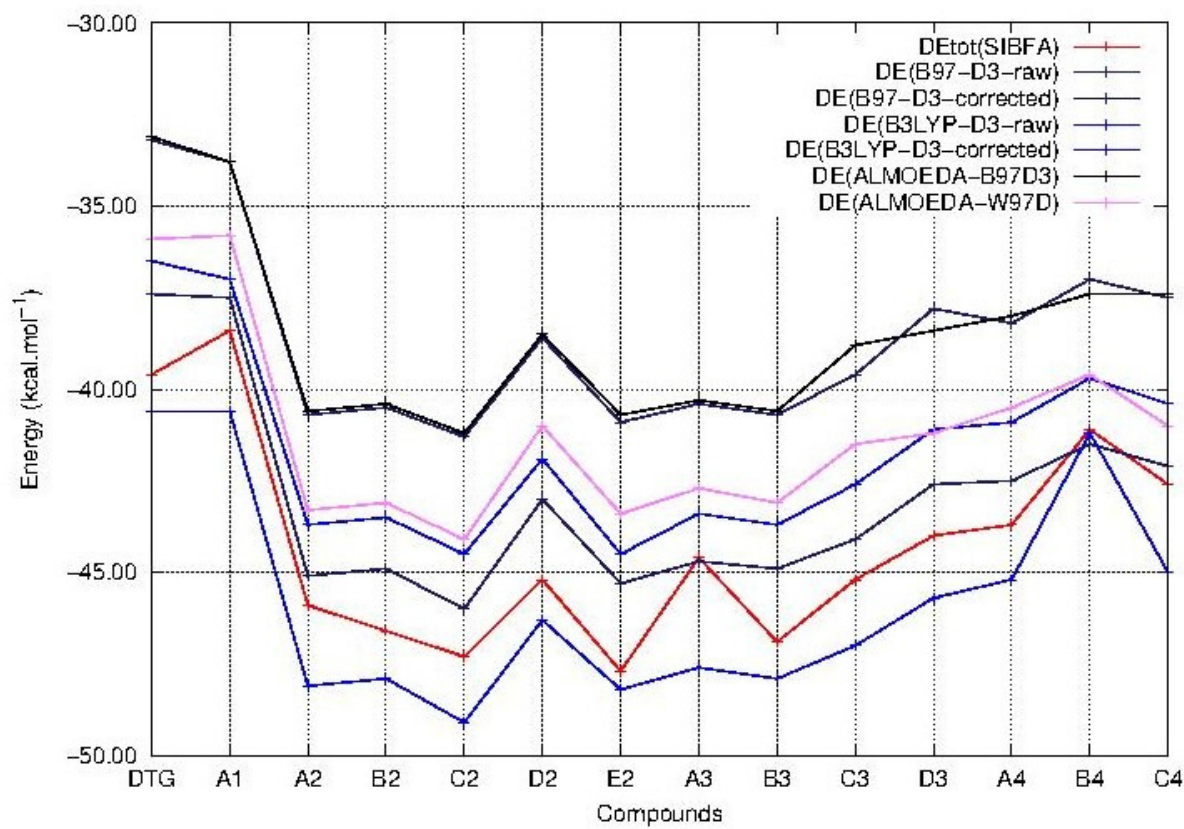

**Figure 4**. Evolution of the intermolecular interaction energies (ΔE) of the ternary complex; $G_4/C_{16}$/substituted ring as a function of the halobenzene ring of DTG and the newly conceived rings (Compounds A1 to C4). ΔE values are calculated via quantum chemistry correlated calculations with the B3LYP-D3 functional (ΔE B3LYP-D3-raw before BSSE correction and ΔE B3LYP-D3-corrected after BSSE correction) and the B97-D3 functional (ΔE B97-D3-raw before BSSE correction and ΔE B97-D3-corrected after BSSE correction). ΔE obtained via quantum chemistry energy decomposition analyses methods (EDA) are also presented, precisely the Absolutely Localized Molecular Orbitals method using two functionals; B97D3 (ΔE ALMOEDA-B97D3) and ω-B97D (ΔE ALMOEDA/ω-B97D). ΔE values are obtained as well via molecular mechanics calculations with the polarizable force field SIBFA (ΔE$_{tot}$ SIBFA)

In figures 5, 6, 7 and 8 we report the evolution of the individual QC contributions along the series. We denote by $E_1$ the 'frozen' energy contribution from ALMOEDA and the sum of the Coulomb, $E_C$, and exchange, $E_X$, contributions from the RVS analyses. $E_1$ (SIBFA) is correspondingly the sum of its penetration-augmented electrostatic and short-range repulsion contributions. We denote by $E_2$ the sum of the polarization and charge-transfer contributions in all approaches.

In the perspective of long-duration polarizable MD on the complexes of vDNA and of the Integrase-vDNA assembly, it is essential to evaluate how well could a procedure such as SIBFA account for the ∆E (QC) trends. This alone could lend credence to comparative energy balances and prospective free energy calculations bearing on complexes out of reach of QC calculations, but amenable to this, and related procedures such as those between 'improved' DTG derivatives with v-DNA, let alone with INT, which total several thousands of atoms. Table 1 and Figure 4 show that $\Delta E_{tot}$ (SIBFA) is fully able to recover the trends from ∆E(QC). C2 and E2 are found to be the two best-bound compounds, while B2, B3, and A2 come next with a small margin. Such agreements are encouraging, considering that no extra calibration effort was done on the DTG derivatives.

Figure 5 reports the evolutions of $E_1$ at the uncorrelated RVS and correlated B97D3 and ω-B97D levels and those of $E_1$ (SIBFA). Figure 6 reports the corresponding evolutions of $E_2$. Figure 7 compares the evolution of uncorrelated ∆E (RVS) and ∆E (SIBFA) without the dispersion contribution $E_{disp}$. Figure 8 reports the evolution of a contribution denoted as '$E_{disp}/E_{corr}$', the gain in ∆E upon passing from the uncorrelated RVS ∆E to the correlated B3LYP-D3, B97-D3, and ω–B97D levels, along with $E_{disp}$ (SIBFA). There is no explicit ALMOEDA 'dispersion' contribution, as it is included in the van der Waals kernel for both $E_1$ and $E_2$.

Throughout we will denote by ∆E (SIBFA) and $\Delta E_{tot}$ (SIBFA) the SIBFA intermolecular interaction energies without and with the dispersion contribution. ∆E (RVS) denotes the BSSE-corrected RVS intermolecular interaction energy. ∆E (QC) denotes the correlated QC intermolecular interaction energy with the B3LYP-D3, B97-D3, or ω–B97D functionals. The QC-derived 'dispersion' energy is the difference between ∆E (QC) and ∆E (HF), the latter being derived from a Hartree-Fock computation without removing the BSSE.

Figure 5 shows that the trends in ΔE (QC) and ΔE$_{tot}$ (SIBFA) giving distinct preferences for derivatives A2, B2, C2, E2, and B3 are retrieved by E$_1$, while the E$_2$ curves (Figure 6) are much shallower. Figure 7 shows that similar to ΔE (QC) and ΔE$_{tot}$ (SIBFA), both ΔE (RVS) and ΔE (SIBFA) curves have minima with derivatives A2-C2, E2 and B3. ΔE (RVS) has an additional minimum with derivative A4 from series 4. The corresponding ΔE (SIBFA) minimum is higher in energy. There is a lesser correspondence between E$_{disp}$ (SIBFA) and those found from the B97D3 and B3LYP-D3 calculations (Figure 8). For compounds A1-E2, it has values intermediate between those from both functionals. For A3 and B3, it is close to the B97D3 values, with differences of 1.2 and 0.2 kcal/mol, respectively. For C3 and D3, it is close to the B3LYP-D3 values, with differences of 0.0 and 0.4 kcal/mol, respectively. For all three compounds in series 4 having the *para* –NH$_2$ substituent, E$_{disp}$ (SIBFA) has smaller values than either functional. In fact, the trends are not consistent between the two functionals. Thus, while E$_{disp}$ (B3LYP-D3) is larger by 2.7 kcal/mol than E$_{disp}$ (B97-D3), both functionals give very close E$_{disp}$ values in both B3 and C3 complexes.

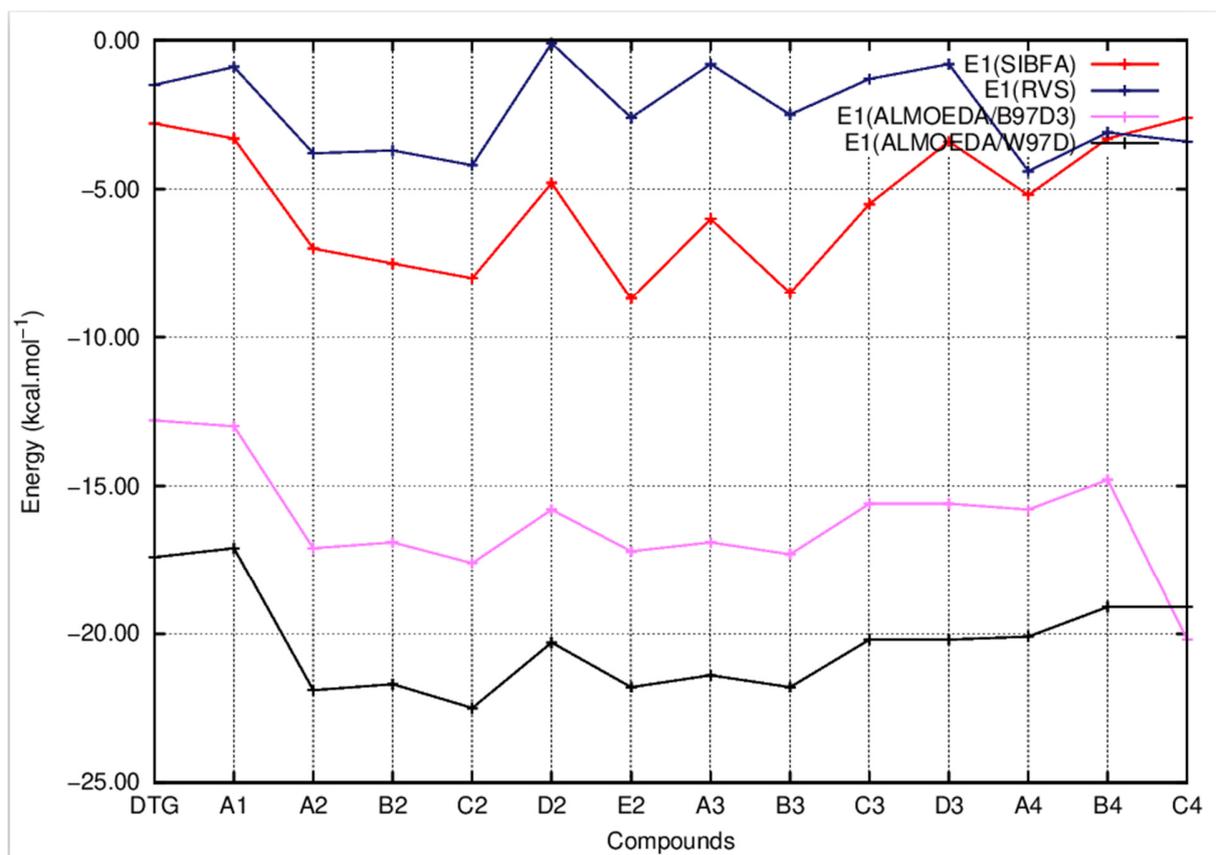

**Figure 5.** Evolution of the first order energy ($E_1$) of the ternary complex; $G_4$/$C_{16}$/substituted ring as a function of the halobenzene ring of DTG and the newly conceived rings (Compounds A1 to C4). $E_1$ values are calculated via quantum chemistry Energy Decomposition Analysis methods (EDA) precisely the Reduced Variational Space method ($E_1$ RVS) and Absolutely Localized Molecular Orbitals method using two functionals; B97D3 ($E_1$ ALMOEDA/B97D3) and ω97D ($E_1$ ALMOEDA/ω-B97D). E1 values are obtained as well via molecular mechanics calculations with the polarizable force field SIBFA ($E_1$ SIBFA).

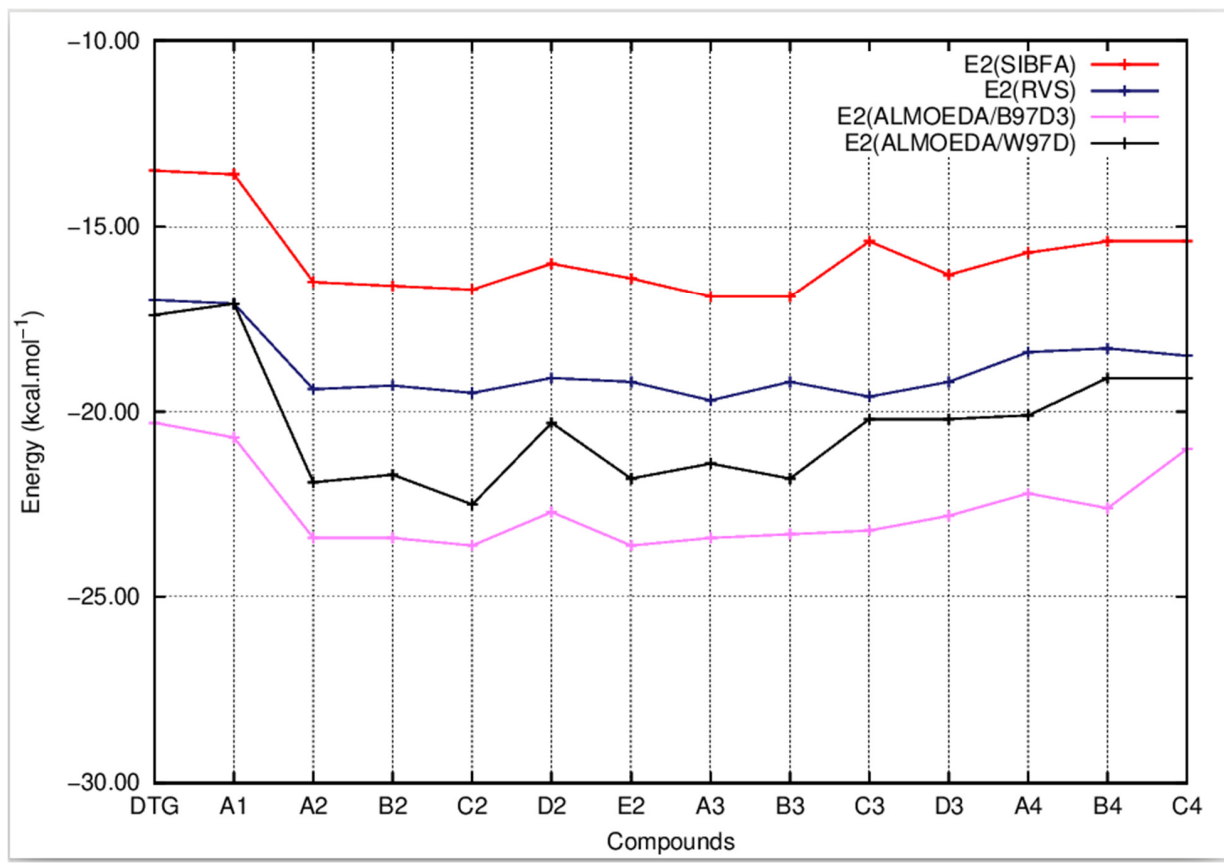

**Figure 6.** Evolution of the second order energy ($E_2$) of the ternary complex; $G_4/C_{16}$/substituted ring as a function of the halobenzene ring of DTG and the newly conceived rings (Compounds A1 to C4). $E_2$ values are calculated via quantum chemistry Energy Decomposition Analysis methods (EDA), precisely the Reduced Variational Space method ($E_2$ RVS) and Absolutely Localized Molecular Orbitals method using two functionals; B97D3 ($E_2$ ALMOEDA/B97D3) and ω-B97D ($E_2$ ALMOEDA/ω-B97D). $E_2$ values were obtained as well via molecular mechanics calculations with the polarizable force field SIBFA ($E_2$ SIBFA).

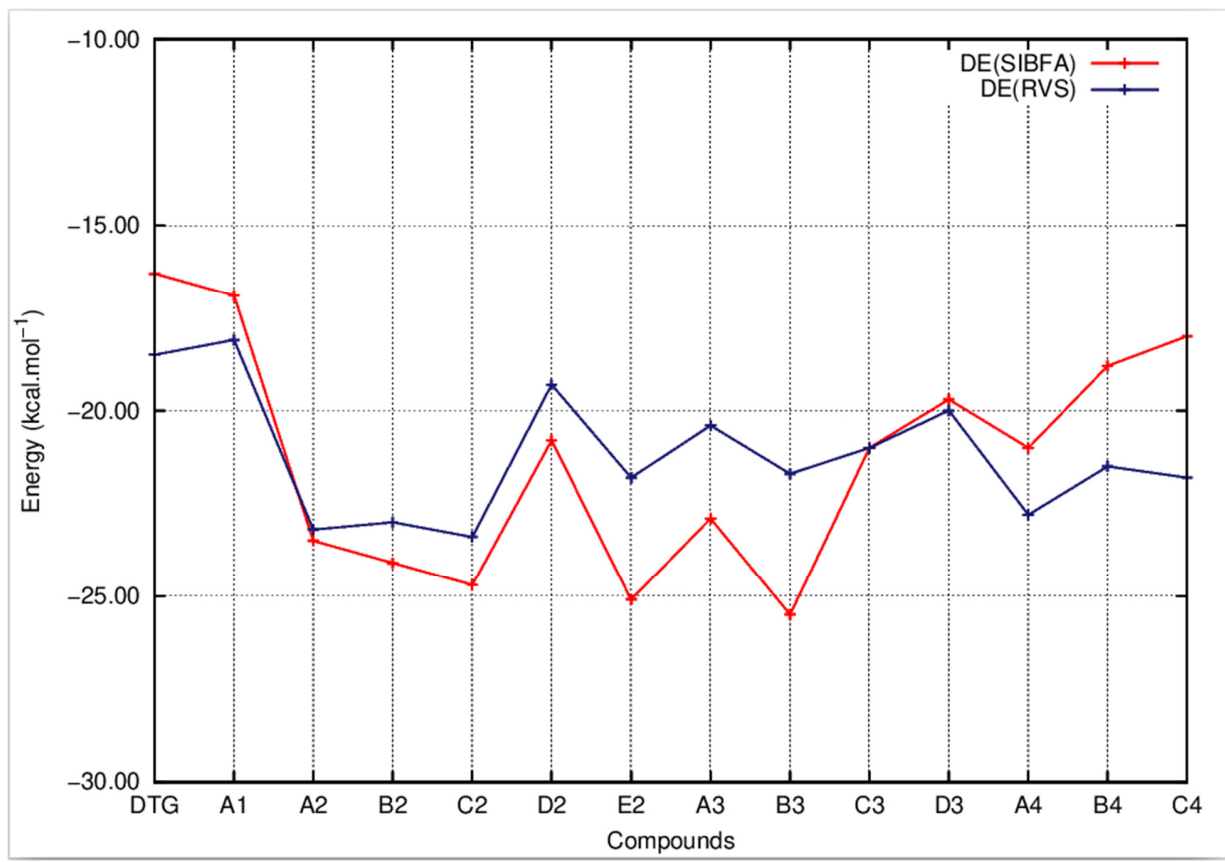

**Figure 7.** Evolution of the interaction energy of the ternary complex; $G_4/C_{16}$/substituted ring as a function of the halobenzene ring of DTG and the newly conceived rings (Compounds A1 to C4). ΔE values are calculated via quantum chemistry Energy Decomposition Analysis method (EDA), precisely the Reduced Variational Space method (ΔE RVS) and via molecular mechanics calculations with the polarizable force field SIBFA (ΔE SIBFA).

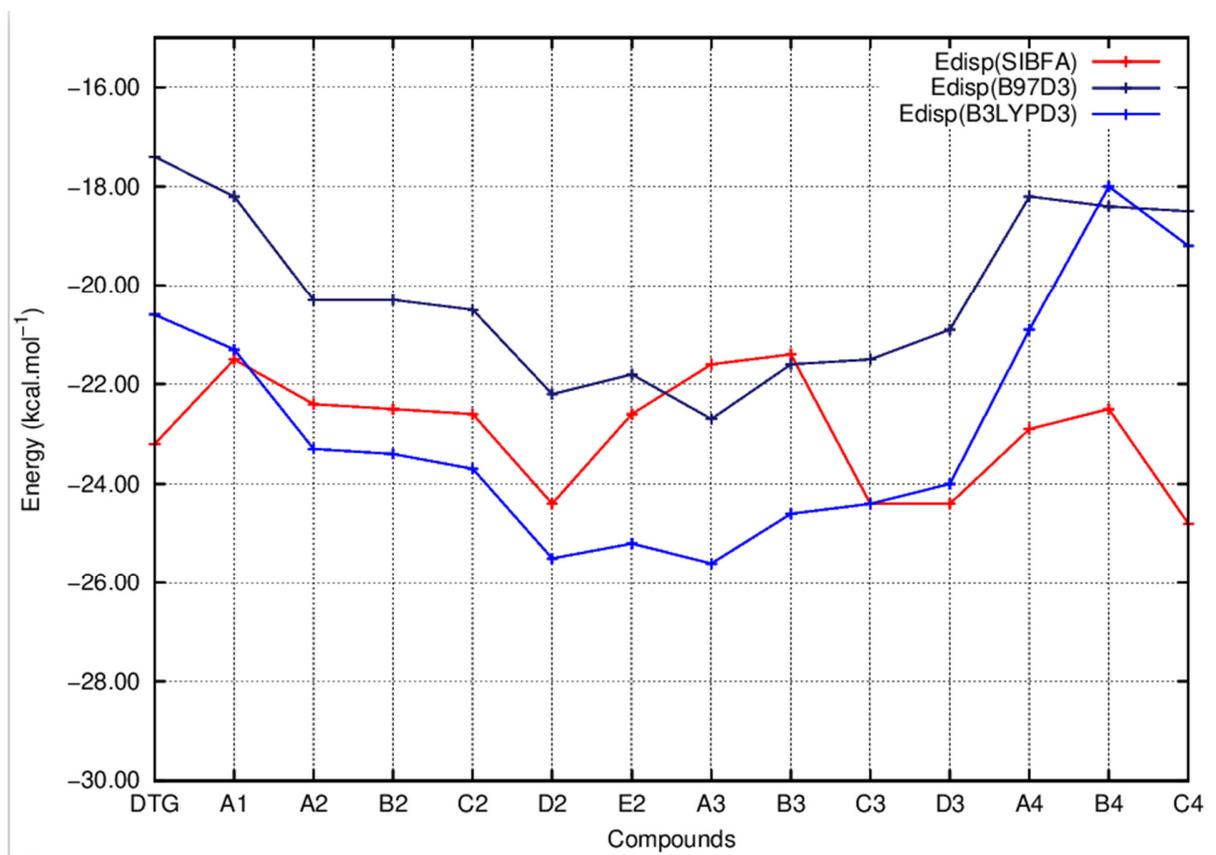

**Figure 8.** Evolution of the dispersion energy ($E_{disp}$) of the ternary complex; $G_4/C_{16}$/substituted ring as a function of the halobenzene ring of DTG and the newly conceived rings (Compounds A1 to C4). $E_{disp}$ values correspond to the ones calculated via quantum chemistry correlated calculations using two functionals; B97D3 ($E_{disp}$ B97D3) and B3LYPD3 ($E_{disp}$ B3LYPD3) and via molecular mechanics calculations with the polarizable force field SIBFA ($E_{disp}$ SIBFA).

Even though it appears satisfactory, the agreement between $\Delta E_{tot}$ (SIBFA) and $\Delta E$ (QC) could be further improved in the near future. There are ongoing SIBFA refinements in the context of reconstruction of a new library of protein, DNA, and ligand constitutive fragments. They resort to multipoles and polarizabilities from correlated calculations, and a rescaling of the individual energy contributions on the basis of correlated Symmetry Adapted Perturbation Theory SAPT-DFT (Podeszwa et al., 2006; Misquitta et al., 2002) energy decomposition analyses. They will be reported elsewhere.

**Molecular Electrostatic Potentials (MEP).**

It is instructive to represent the impact of some of the reported substitutions on the MEP contours. Figure 9 presents such contours around five representative compounds: DTG; C2 and E2; the *meta*-substituted trifluoro- and cyano derivatives of group 2, respectively; B3, the *meta*-substituted bromine derivative of series 3; and C4, the difluorine-substituted derivative of group 4.

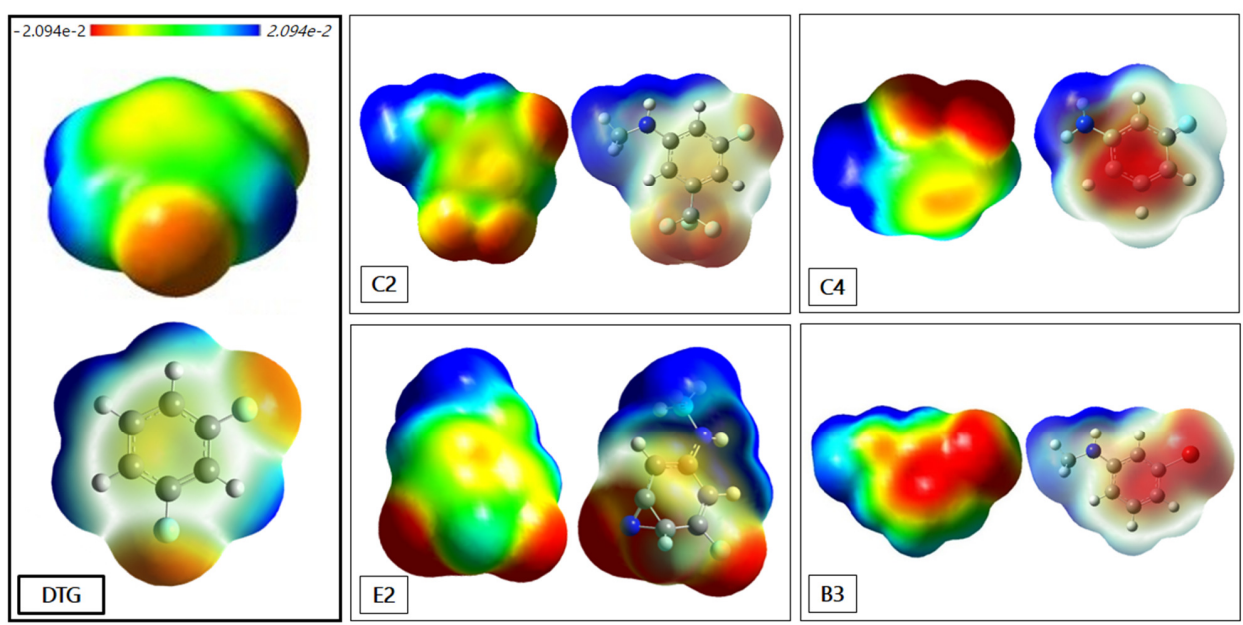

**Figure 9.** Contours of the Molecular Electrostatic Potential (MEP) around: DTG, C2, E2, B3 and C4. The distribution of the electronic charges is $10^{-3}$ electron Bohr-3 isodensity surface. The colored bar at the top of the figure indicates the variation of the electron density between $-2.094^{e-2}$ (corresponding to the color red) and $2.094^{e-2}$ (corresponding to the color blue)

There is a much wider extension of the 'blue' zone of positive potential around the *para* – $NHCH_3$ or $–NH_2$ substituents than around the *para*-fluorine substituent of DTG. This explains the better efficiency observed in the case of the newly conceived compounds in comparison with DTG. In fact, the lower electron density of $NHCH_3$ and $NH_2$ confirms the better electrostatic interaction of these rings with the electron rich polycyclic structure of $G_4$. For the compounds of group 2, the $–CF_3$ substituted derivative C2 generates a much wider and delocalized 'red' zone of negative potential than the cyano-substituted derivative E2. The extension of positive potential around the -NHCH$_3$ group is also larger for C2 than for E2. To a large extent, the first feature should explain the 2.1 kcal/mol larger $\Delta G_{solv}$ energy of C2 than E2, impacting the energy balance in favor of E2 while both derivatives had close $\Delta E$ (QC) and $\Delta E_{tot}$ (SIBFA) values. Derivative B3 also displays a wide zone of negative electrostatic potential around the –Br substituent, and this zone is significantly more extended than around the two –F substituents of derivative C4. The latter presents a low electron density in the middle of its ring due to the presence of its two-electron attracting fluorines, giving it the least favorable attraction with the DNA viral base pairs

**Conclusions and Perspectives.**

We have designed a series of derivatives of the halobenzene ring of dolutegravir (DTG), the most potent inhibitor of the HIV-1 integrase to date. These derivatives target the highly conserved $G_4/C_{16}$ base pair of viral DNA. The -*para* fluorine ring of DTG is replaced by a –NHCH$_3$ or a –NH$_2$ substituent. All derivatives reported in this study had more favorable intermolecular interaction energies with this base pair than DTG, as computed by quantum chemistry, and a polarizable molecular mechanics procedure, SIBFA. The two compounds with the highest $G_4/C_{16}$ affinities had a *para* –NHCH$_3$ substituent and two substituents *meta* to it: fluorine and trifluromethyl, denoted as C2, and fluorine and cyano, denoted as E2. $\Delta E_{tot}$ (SIBFA) displayed trends consistent with $\Delta E$ (QC). This study, along with our previous ones (El Hage et al., 2014, 2015; El Khoury et al., 2019) shows that it is possible to design, in a piece-wise fashion, novel derivatives targeting a well-defined, highly conserved, subset of the recognition site of a large macromolecular target. It also constitutes an essential validation step prior to large-scale polarizable molecular dynamics simulations of the complex of the entirety of the drug with the entirety of the target.

**Acknowledgments.**

This work has received funding from the European Research Council (ERC) under the European Union's Horizon 2020 research and innovation program (grant agreement No 810367), project EMC2. We wish to thank the Grand Equipement National de Calcul Intensif (GENCI): Institut du Développement et des Ressources en Informatique Scientifique (IDRIS), Centre Informatique de l'Enseignement Supérieur (CINES), France, project No. x2009-075009), and the Centre Régional Informatique et d'Applications Numériques de Normandie, Rouen, France), project 1998053. We thank also for funding this work the Research Council of Saint-Joseph University of Beirut, Lebanon (Project FS71), the Lebanese National Council for Scientific Research, CNRS-L (Projects CNRS-FS80 and CNRS-FS 116), the French Institute- Lebanon and the French-Lebanese Program CEDRE Project: 35327UJ.

**Figure captions.**

**Figure 1.** Molecular structure of Raltegravir (RAL), Elvitegravir (EVG) and Dolutegravir (DTG) represented with the Chimera software. The coordinates used for these representations are taken from the Protein Data Bank crystal structures of the Prototype Foamy Virus (PFV) intasome in complex with magnesium and Raltegravir (PDB code: 3OYA - Hare et al., 2010a), Elvitegravir (PDB code: 3L2U, Hare et al., 2010b) as well as with Dolutegravir (PBD code: 3S3M - Hare et al., 2011). We emphasized with a circle the difluoro halobenzyl ring of DTG, where the substitutions are going to be made to conceive new inhibitors. Carbon atoms are colored tan, hydrogen atoms are colored white, fluorine atoms are colored light green, chlorine atoms are colored dark green, nitrogen atoms are colored blue and oxygen atoms are colored red.

**Figure 2.** Molecular structures of the halobenzene ring of DTG (Group 1) and the newly designed ones (Groups 1, 2, 3 and 4) represented with the Chimera software. Carbon atoms are colored tan, hydrogen atoms are colored white, fluorine atoms are colored light green, chlorine atoms are colored dark green, bromine atoms are colored dark red and nitrogen atoms are colored blue. The squares indicate the new atoms or moieties chosen after substitution

**Figure 3.** Representation of the $G_4/C_{16}$ complexes with representative HRs. The structures of $G_4$, $C_{16}$ and DTG are taken from the Protein Data Bank crystal structure of the Prototype Foamy Virus (PFV) intasome in complex with magnesium and Dolutegravir (PDB code: 3S3M Hare et al., 2011). The newly designed compounds are represented with the Chimera software. The color code for the atoms remains the same as in Figure 1, with the oxygen atoms colored in ligh red.

**Figure 4**. Evolution of the intermolecular interaction energies ($\Delta E$) of the ternary complex; $G_4/C_{16}$/substituted ring as a function of the halobenzene ring of DTG and the newly conceived rings (Compounds A1 to C4). $\Delta E$ values are calculated via quantum chemistry correlated calculations with the B3LYP-D3 functional ($\Delta E$ B3LYP-D3-raw before BSSE correction and $\Delta E$ B3LYP-D3-corrected after BSSE correction) and the B97-D3 functional ($\Delta E$ B97-D3-raw before BSSE correction and $\Delta E$ B97-D3-corrected after BSSE correction). $\Delta E$ obtained via quantum chemistry energy decomposition analyses methods (EDA) are also presented, precisely the Absolutely Localized Molecular Orbitals method using two functionals; B97D3 ($\Delta E$ ALMOEDA-B97D3) and ω-B97D ($\Delta E$ ALMOEDA/ω-B97D). $\Delta E$ values are obtained as well via molecular mechanics calculations with the polarizable force field SIBFA ($\Delta E_{tot}$ SIBFA)

**Figure 5.** Evolution of the first order energy ($E_1$) of the ternary complex; $G_4/C_{16}$/substituted ring as a function of the halobenzene ring of DTG and the newly conceived rings (Compounds A1 to C4). $E_1$ values are calculated via quantum chemistry Energy Decomposition Analysis methods (EDA) precisely the Reduced Variational Space method ($E_1$ RVS) and Absolutely Localized Molecular Orbitals method using two functionals; B97D3 ($E_1$ ALMOEDA/B97D3) and ω97D ($E_1$ ALMOEDA/ω-B97D). E1 values are obtained as well via molecular mechanics calculations with the polarizable force field SIBFA ($E_1$ SIBFA).

**Figure 6.** Evolution of the second order energy ($E_2$) of the ternary complex; $G_4/C_{16}$/substituted ring as a function of the halobenzene ring of DTG and the newly conceived rings (Compounds A1 to C4). $E_2$ values are calculated via quantum chemistry Energy Decomposition Analysis methods (EDA), precisely the Reduced Variational Space method ($E_2$ RVS) and Absolutely Localized Molecular Orbitals method using two functionals; B97D3 (E2 ALMOEDA/B97D3) and ω-B97D (E2 ALMOEDA/ω-B97D). $E_2$ values were obtained as well via molecular mechanics calculations with the polarizable force field SIBFA ($E_2$ SIBFA).

**Figure 7.** Evolution of the interaction energy of the ternary complex; $G_4/C_{16}$/substituted ring as a function of the halobenzene ring of DTG and the newly conceived rings (Compounds A1 to C4). ΔE values are calculated via quantum chemistry Energy Decomposition Analysis method (EDA), precisely the Reduced Variational Space method (ΔE RVS) and via molecular mechanics calculations with the polarizable force field SIBFA (ΔE SIBFA).

**Figure 8.** Evolution of the dispersion energy ($E_{disp}$) of the ternary complex; $G_4/C_{16}$/substituted ring as a function of the halobenzene ring of DTG and the newly conceived rings (Compounds A1 to C4). $E_{disp}$ values correspond to the ones calculated via quantum chemistry correlated calculations using two functionals; B97D3 ($E_{disp}$ B97D3) and B3LYPD3 ($E_{disp}$ B3LYPD3) and via molecular mechanics calculations with the polarizable force field SIBFA ($E_{disp}$ SIBFA).

**Figure 9.** Contours of the Molecular Electrostatic Potential (MEP) around: DTG, C2, E2, B3 and C4. The distribution of the electronic charges is $10^{-3}$ electron Bohr-3 isodensity surface. The colored bar at the top of the figure indicates the variation of the electron density between $-2.094^{e-2}$ (corresponding to the color red) and $2.094^{e-2}$ (corresponding to the color blue)

**Table 1.** Intermolecular interaction energies (kcal/mol) of halogenated derivatives with the $G_4/C_{16}$ vDNA base pair. DTG as well as the compounds displaying the best energetic values are highlighted in red.